\documentclass[a4paper,12pt]{article}
\usepackage{graphics}
\usepackage{amsfonts}

\begin{document}
%

\newcommand{\be}{\begin{equation}}
\newcommand{\ee}{\end{equation}}
\newcommand{\bea}{\begin{eqnarray}}
\newcommand{\eea}{\end{eqnarray}}
\newcommand{\bean}{\begin{eqnarray*}}
\newcommand{\eean}{\end{eqnarray*}}
\font\upright=cmu10 scaled\magstep1
\font\sans=cmss12
\newcommand{\ssf}{\sans}
\newcommand{\stroke}{\vrule height8pt width0.4pt depth-0.1pt}
\newcommand{\Z}{\hbox{\upright\rlap{\ssf Z}\kern 2.7pt {\ssf Z}}}
\newcommand{\ZZ}{\Z\hskip -10pt \Z_2}
\newcommand{\C}{{\rlap{\upright\rlap{C}\kern 3.8pt\stroke}\phantom{C}}}
\newcommand{\R}{\hbox{\upright\rlap{I}\kern 1.7pt R}}
\newcommand{\HH}{\hbox{\upright\rlap{I}\kern 1.7pt H}}
\newcommand{\CP}{\hbox{\C{\upright\rlap{I}\kern 1.5pt P}}}
\newcommand{\identity}{{\upright\rlap{1}\kern 2.0pt 1}}
\newcommand{\half}{\frac{1}{2}}
\newcommand{\quart}{\frac{1}{4}}
\newcommand{\pr}{\partial}
\newcommand{\bm}{\boldmath}
\newcommand{\I}{{\cal I}} 
\newcommand{\M}{{\cal M}}
\newcommand{\N}{{\cal N}}
\newcommand{\e}{\varepsilon}
\newcommand{\ep}{\epsilon}
\newcommand{\bep}{\mbox{\boldmath $\varepsilon$}}
\newcommand{\Oh}{{\rm O}}
\newcommand{\x}{{\bf x}}
\newcommand{\y}{{\bf y}}
\newcommand{\X}{{\bf X}}
\newcommand{\Y}{{\bf Y}}
\newcommand{\z}{{\bar z}}
\newcommand{\w}{{\bar w}}
\newcommand{\tT}{{\tilde T}}
\newcommand{\tX}{{\tilde\X}}

\thispagestyle{empty}
\rightline{DAMTP-2013-70}
\vskip 3em
\begin{center}
{{\bf \Large Newtonian Atlas for Dust-Filled FRW Universe
}} 
\\[15mm]

{\bf \large N.~S. Manton\footnote{email: N.S.Manton@damtp.cam.ac.uk}} \\[20pt]

\vskip 1em
{\it 
Department of Applied Mathematics and Theoretical Physics,\\
University of Cambridge, \\
Wilberforce Road, Cambridge CB3 0WA, U.K.}
\vspace{12mm}

\abstract{}
The metric of an FRW universe filled with pressureless dust is shown
to agree, close to any spacetime point, with a curved Newtonian-type
metric where Einstein's equations simplify to those of Newtonian
gravity. The agreement is shown to quadratic order in the local
coordinates, so the curvatures agree. This result is established by 
expressing both metrics in Riemann normal form. This
approach gives a local Newtonian understanding of cosmology that
avoids the paradoxes of global Newtonian cosmology.

\end{center}

\vskip 100pt
\leftline{Keywords: FRW Universe, Newtonian Atlas, 
Riemann Curvature, Normal} 
\leftline{Coordinates}
\vskip 5pt

\vfill
\newpage
\setcounter{page}{1}
\renewcommand{\thefootnote}{\arabic{footnote}}


\section{Introduction} 
\vspace{4mm}

It is well known that the evolution equations of a
Friedmann--Robertson--Walker (FRW) universe \cite{Wei} are the same as those
obtained in Newtonian cosmology \cite{Mil,McM,Bon,Cal}. In particular, 
in the most Newtonian case, that of pressureless dust with no 
cosmological constant, the evolution can be interpreted as due to
the Newtonian self-gravity of the dust. The FRW and Newtonian geometries are
however globally different. The FRW universe has spacetime curvature, and
the cross sections at fixed comoving time $t$ are generically 
spatially curved. If the spatial sections have positive curvature,
they are of finite size. The Newtonian universe is flat and infinite. 

The Newtonian universe can consist of a bounded ball of matter,
surrounded by an infinite empty region \cite{McC}, and the matter may be 
discrete \cite{GE}. Alternatively it can consist
of a uniform distribution of matter over infinite space. Both models 
have disturbing features \cite{Lay,Nor}. In the first case there is a
definite centre, so a lack of homogeneity except near the centre. In
the second case, one problem is that the gravitational 
potential $\Phi$ of uniform matter is indeterminate if calculated by 
the inverse square law. Also the Hubble flow, where the matter
velocity away from a point is proportional to the distance from the
point, has the property that beyond a certain distance the velocity is 
greater than the speed of light. The first problem can be resolved 
by solving Poisson's equation for the potential $\Phi$. There is 
a spherically symmetric solution around a given point which
grows quadratically with distance. This potential is also obtained if
one imagines removing a ball of matter around the point, arguing 
that the potential inside is then constant, and then putting the 
ball back. The solution has the paradoxical property of again 
having a definite centre despite the homogeneity of the background 
matter density, and it leads to the acceleration of matter being towards this
centre. The second problem is less severe if one 
regards the speed of light as irrelevant in Newtonian dynamics.  

We will show that the Newtonian view is more accurate and 
devoid of paradox if one regards the FRW universe as covered by an 
atlas of overlapping patches. This develops the idea that
Newtonian dynamics is effective in small regions of the
universe, see e.g. \cite{Har, Cal}. We will examine in detail the
geometry of these patches and how neighbouring patches fit
together. Our patches are of finite size but do not have a sharp
physical cutoff. The cutoff will be purely mathematical, as in 
differential geometry, where one covers a manifold
with an atlas of overlapping charts.

Around each point $\Oh$ of an FRW spacetime we establish that there is
a Newtonian approximation up to quadratic order in the Riemann normal
coordinates centred at $\Oh$. At this order one captures
the Riemann curvature, so the Newtonian patches are not just tangent
planes, but osculate the spacetime manifold.  

By Newtonian patch, with a Newtonian-type metric, we mean the curved 
spacetime that is regarded as capturing the Newtonian limit of general
relativity \cite{Wei,Car}. It is a solution of Einstein's equations 
in the weak field limit, close to flat spacetime, with uniform dust as 
source. The solution depends on the local Newtonian potential 
$\Phi$, and the geodesic motion of non-relativistic test particles 
as well as the dust itself experiences an acceleration proportional 
to the gradient of $\Phi$. It is sufficient to work with the metric 
whose source is dust at rest. The Hubble flow away from or towards 
the central point (both its velocity and acceleration) produces 
corrections at higher order.

Technically, we will show that by coordinate
transformations around $\Oh$, we can put the
FRW metric into Riemann normal form up to quadratic order in the
coordinates (with the coordinates vanishing at $\Oh$). We will
also put the Newtonian-type metric into Riemann normal form, and
show that the FRW and Newtonian-type metrics are then the same. 
The Hubble flows also agree at linear order. The Riemann normal form 
is where the metric to leading order is Minkowskian, all the first
derivatives of the metric vanish, so the Christoffel symbols
vanish at $\Oh$, and additionally the second derivatives of 
the metric are directly related to the Riemann curvature \cite{Spi}. 
It occurs when the coordinates are geodesic normal coordinates to
quadratic order. To find it, we will proceed by algebraic trial 
and error, and not actually construct 
geodesics emanating from $\Oh$. This method is quite simple because 
of the isotropy of the metric around $\Oh$.

It is interesting to keep track of the number of parameters that occur
in the metrics, at quadratic order. The Robertson--Walker metric,
expanded about $\Oh$, has three continuous parameters, the Hubble 
parameter $H$, the deceleration $q$, and the spatial curvature 
$K = k/R_0^2$. $k$ here has its usual discrete values,
$1,0$ or $-1$. However, the Riemann curvature has just two parameters,
a spatial part and a time-space part, which are combinations of the
three just mentioned. Then the Friedmann equations (the Einstein equations) 
relate both curvature parameters to the matter density $\rho$ at
$\Oh$. (The matter density in comoving coordinates is a Lorentz scalar.) The
metric in Riemann normal form therefore depends on just one quantity, 
the density $\rho$.

Exactly the same final parameter count occurs in the Newtonian patches.  
The metric is determined by the potential $\Phi$, which in turn 
is determined by the density. At quadratic order there is no further 
information. 

This is rather curious. It means that from the spacetime curvature
alone, at one point, one cannot determine all of
$K$, $q$ and $H$, but only two combinations of these. However, the
Hubble parameter $H$ is separately determined from the 
flow of matter away from $\Oh$, and is measured using redshifts. In 
turn, the Hubble flow controls the rate of change of the 
density $\rho$ and the form of the energy-momentum tensor 
in the neighbourhood of $\Oh$. In detail, the energy-momentum tensor 
depends on the coordinate system, so it is different in comoving 
coordinates and Riemann normal coordinates. 

It is now fashionable to interpret the redshift of photons as due 
to the spatial expansion of the whole universe \cite{Wei,Har}. In an 
FRW universe, and calculating over long time intervals, it is clearly 
convenient to use comoving coordinates, and the well known result that 
redshift $z$ depends only on the ratio of the scale 
factors at the times of emission and receipt of photons is
mathematically elegant. 

However, space doesn't really expand, at least not locally. Near $\Oh$, 
one just has a Newtonian patch of a Lorentzian curved spacetime, whose metric 
one can approximate using Riemann normal coordinates. Over short 
distances and times even the curvature can be ignored, so spacetime 
is approximately Minkowskian. In these coordinates there is a Hubble 
flow of light-emitting matter, and the redshift is a Doppler effect 
due to the recession of the emitters from $\Oh$.

This remark is conceptual rather than practical. The more sophisticated
point of view, that redshift is due to the expansion of space 
is helpful, but it relies on the global form and symmetries of the FRW
metric and on the use of comoving coordinates. Locally, 
in normal coordinates, redshift is a Doppler effect. The use of Newtonian 
patches justifies this traditional understanding of redshift.  

\vspace{7mm}

\section{FRW metrics in Riemann normal form}
\vspace{4mm}

Here we show, by suitable coordinate transformations, that an FRW
metric, expanded around a spacetime point $\Oh$, can be brought to 
Riemann normal form.

We recall that a Robertson--Walker metric is of the form
\be
ds^2 = - dt^2 + R(t)^2\frac{d{\x'} \cdot d{\x'}}
{\left(1 + \quart k {\x'} \cdot {\x'} \right)^2}
\ee
where $k$ is $1,0$ or $-1$. $t$ and ${\x'}$ are comoving coordinates. 
The spatial metric here, the expression multiplied by $R(t)^2$, is one 
way of writing a metric of constant curvature $k$. Without loss of generality, 
we choose $\Oh$ to be at $\x' ={\bf 0}$ and (following a shift of $t$ if
necessary) at $t=0$. The standard way to write the Taylor expansion of 
the scale factor $R(t)$ around $t=0$, up to quadratic order, is
\be
R(t) = R_0\left(1 + Ht - \half q H^2 t^2 \right) \,.  
\label{scalefac}
\ee
$H$ is the Hubble parameter and $q$ the deceleration at $t=0$. 
[If $H$ is zero, one needs to write $R(t) = R_0(1 - \half Q t^2)$.] 
Defining $\x = R_0 {\x'}$, the metric becomes
\be
ds^2 = - dt^2 + \left(1 +2Ht + (1 - q)H^2 t^2 \right)
\frac{d\x \cdot d\x}{\left(1 + \frac{k}{4R_0^2} \x \cdot \x \right)^2} \,,
\ee
and further expanding about $\Oh$, i.e. about $\x = {\bf 0}$, we find 
to quadratic order
\be
ds^2 = - dt^2 + \left(1 +2Ht + (1 - q)H^2 t^2 - \half K\x \cdot \x
\right) d\x \cdot d\x \,.
\label{FRWquad}
\ee
$K = k/R_0^2$ is the curvature of the spatial cross section at $t=0$, with the
scale factor taken into account, and it takes any real value.
Our analysis is formal throughout. We do not draw attention to higher 
order terms, and in all expressions it is implied that there are 
higher order corrections that have been dropped.

From the Einstein equations, one obtains the Friedmann equations for
the scale factor $R(t)$. In the simplest case that the matter is
pressureless dust of density $\rho(t)$, and there is no cosmological
constant, the Friedmann equations are
\be
\frac{\ddot R}{R} = -\frac{4\pi G}{3}\rho
\label{Friedeq}
\ee
and
\be
\frac{{\dot R}^2}{R^2} + \frac{k}{R^2} = \frac{8\pi G}{3}\rho \,,
\label{Raych}
\ee
together with the mass conservation constraint
\be
\rho R^3 = M = {\rm constant} \,. 
\label{masscons}
\ee
These are related, as by eliminating $\rho$ in favour of $M$ 
in (\ref{Raych}), and differentiating, one obtains (\ref{Friedeq}).

In terms of the Taylor coefficients of $R$, as in (\ref{scalefac}),
eqs.(\ref{Friedeq}) and (\ref{Raych}) reduce at $t=0$ to
\bea
qH^2 &=& \frac{4\pi G}{3} \rho_0 \,, \label{Friedq}\\ 
H^2 + K &=& \frac{8\pi G}{3}\rho_0 \,, \label{FriedHK} 
\eea
where $\rho_0 = \rho(0)$. It follows that $K = (2q-1)H^2$. 

The metric (\ref{FRWquad}) is the one we shall now manipulate using
coordinate transformations. It describes the FRW geometry up to
quadratic order in both space and time around $\Oh$. Standard
calculations will correctly determine the Riemann curvature at $\Oh$.
However, by bringing the metric to Riemann normal form, the curvature
will become manifest.

The leading terms in (\ref{FRWquad}) have
Minkowski form, but we need to remove the term linear in $t$. The desired
coordinate change, involving quadratic terms in the coordinates (and
no cubic terms yet), is
\bea
t &=& \tT - \half H \, \tX \cdot \tX \,, \nonumber \\
\x &=& \tX - H \, \tT\tX \,,  \label{tX}
\eea
implying
\bea
dt &=& d\tT - H \tX \cdot d\tX \,, \nonumber \\
d\x &=& d\tX  - H (\tX \, d\tT + \tT \, d\tX) \,.
\eea
One finds, keeping all terms at quadratic order,
\bea
ds^2 &=& - d\tT^2 + d\tX \cdot d\tX \nonumber \\
&& + H^2 \, \tX \cdot \tX \, d\tT^2 - 2H^2 \, \tX \cdot d\tX \, \tT d\tT 
- (2+q)H^2 \, \tT^2 \, d\tX \cdot d\tX \nonumber \\
&& - H^2(\tX \cdot d\tX)^2 - 
\left(H^2 + \half K \right)\tX \cdot \tX \, d\tX \cdot d\tX \,.
\label{metricquad}
\eea
The Minkowskian leading terms and absence of linear terms imply that the
coordinates $\tT , \tX$ are normal coordinates to lowest order, and
that the Christoffel symbols at $\Oh$ vanish. To put the metric 
in Riemann normal form at quadratic order, we need a further 
coordinate transformation, involving cubic terms, so that 
the quadratic terms in the metric can be expressed as a quadratic 
form in the antisymmetrised quantities $X^\mu dX^\nu - X^\nu dX^\mu$, 
where $X^\mu$ runs over all four coordinates.

Spherical symmetry around $\Oh$ makes it relatively easy to do this. 
We carry out the coordinate change
\bea
\tT &=& T +  \frac{1}{6}(3+q)H^2 \, T \, \X \cdot \X \,,
\nonumber \\
\tX &=& \X +  \frac{1}{3}(3+q)H^2 \, T^2 \, \X 
+ \frac{1}{6}\left(2H^2 + \half K \right) (\X \cdot \X) \X \,,
\label{TX}
\eea
and hence
\bea
d\tT &=& dT + \frac{1}{6}(3+q)H^2 
(\X \cdot \X \, dT + 2T \, \X \cdot d\X) \,, \nonumber \\
d\tX &=& d\X + \frac{1}{3}(3+q)H^2(2T \, \X \, dT + T^2 \, d\X) \nonumber \\ 
&& + \frac{1}{6}\left(2H^2 + \half K \right)
(2\X (\X \cdot d\X) + (\X \cdot \X) d\X) \,.
\eea
This preserves the spherical symmetry of (\ref{metricquad}). The
coefficients $\frac{1}{6}(3+q)H^2$ etc. are not initially fixed, 
but by simple algebra one can show that with the values shown 
the metric becomes
\bea
ds^2 &=& - dT^2 + d\X \cdot d\X \nonumber \\
&& - \frac{1}{3}qH^2 (\X \, dT - T \, d\X)\cdot(\X \, dT - T \, d\X) 
\nonumber \\ 
&& - \frac{1}{3}(H^2 + K) 
((\X \cdot \X)d\X \cdot d\X - (\X \cdot d\X)^2) \,.
\label{PreRiemNorm}
\eea
The spatial terms $(\X \cdot \X)d\X \cdot d\X 
- (\X \cdot d\X)^2$ can be rewritten as
\be
(X^1 dX^2 - X^2 dX^1)^2 
+ (X^2 dX^3 - X^3 dX^2)^2 + (X^3 dX^1 - X^1 dX^3)^2 \,,
\ee
so the metric (\ref{PreRiemNorm}) has the Riemann normal form 
\bea
ds^2 &=& - dT^2 + d\X \cdot d\X \nonumber \\
&& - \frac{1}{3}qH^2 \Bigl((X^1 dT - T dX^1)^2
+ (X^2 dT - T dX^2)^2 + (X^3 dT - T dX^3)^2 \Bigr) \nonumber \\ 
&& - \frac{1}{3}(H^2 + K) 
\Bigl((X^1 dX^2 - X^2 dX^1)^2 + (X^2 dX^3 - X^3 dX^2)^2 \nonumber \\ 
&& \quad\quad\quad\quad\quad\quad\quad\quad\quad\quad\quad\quad
\quad\quad\quad\quad + (X^3 dX^1 - X^1 dX^3)^2 \Bigr) \,.
\label{RiemNorm}
\eea
(As usual, we identitfy $X^0 = T$.) Spherical symmetry alone allows
the term $(X^1 dT - T dX^1)(X^2 dX^3 - X^3 dX^2)$ and its cyclic 
permutations in the metric, but they are ruled out by 
inversion symmetry, and do not arise under the coordinate change.

For a general metric in Riemann normal form, the quadratic terms 
have the structure 
\be
ds^2_{\rm quad.} = \sum_{\mu,\nu,\sigma,\tau} C_{\mu\nu\sigma\tau} 
(X^\mu dX^\nu - X^\nu dX^\mu) (X^\sigma dX^\tau - X^\tau dX^\sigma) \,,
\ee
where the coefficients $C_{\mu\nu\sigma\tau}$ have the same symmetry properties 
as the Riemann tensor, namely antisymmetry under exchange of $\mu,\nu$ or of
$\sigma,\tau$, symmetry under exchange of $\mu\nu$ with $\sigma\tau$, 
and the cyclic symmetry $C_{\mu\nu\sigma\tau} + C_{\mu\sigma\tau\nu} 
+ C_{\mu\tau\nu\sigma} = 0$. (As explained in \cite{Spi}, this
was the way Riemann initially understood curvature. Note also that our 
index conventions differ from those in \cite{Spi}.) By expanding out, 
and using the standard formula for the Riemann tensor
$R_{\mu\nu\sigma\tau}$, one finds 
$R_{\mu\nu\sigma\tau} = 12 \, C_{\mu\nu\sigma\tau}$.  

For the metric (\ref{RiemNorm}), the consequences of the Einstein 
equations can therefore be verified directly. The non-vanishing 
Riemann tensor components are
\bea
R_{0i0i} &=& -qH^2  \,, \nonumber \\
R_{ijij} &=& -(H^2 + K) \,, \quad\quad (i,j=1,2,3 \,, \,\, i\ne j) \,,
\eea
and the further components implied by the symmetries of the 
Riemann tensor. The Ricci tensor is diagonal, with components $R_{00} =
-3qH^2$ and $R_{ii} = (q-2)H^2 - 2K$ (no sum over $i$), so the 
Einstein tensor has non-vanishing components
\bea
G_{00} &=& -3(H^2 + K) \,, \nonumber \\
G_{ii} &=& -(2q-1)H^2 + K \,.
\eea
The energy-momentum tensor for dust is $T^{\mu\nu} = \rho u^\mu
u^\nu$, where $u^\mu$ is its local 4-velocity. In the comoving frame 
$u^\mu = (1,0,0,0)$ so the only non-vanishing component is 
$T^{tt} = \rho$. Our changes of coordinates do not change the 4-velocity
at $\Oh$, so the only non-vanishing component at $\Oh$ (in Riemann
normal coordinates) becomes $T^{00} = \rho_0$. The Einstein equations 
are therefore $G_{00} = -8\pi G \rho_0$ and $G_{ii} = 0$. These reproduce the 
equations (\ref{Friedq}) and (\ref{FriedHK}).

Notice that eqs.(\ref{Friedq}) and (\ref{FriedHK}) imply that the 
metric (\ref{RiemNorm}) simplifies further, to
\bea
ds^2 &=& - dT^2 + d\X \cdot d\X \nonumber \\
&& - \frac{4\pi G}{9} \rho_0 \Bigl((X^1 dT - T dX^1)^2
+ (X^2 dT - T dX^2)^2 + (X^3 dT - T dX^3)^2 \Bigr) \nonumber \\ 
&& - \frac{8\pi G}{9} \rho_0 \Bigl((X^1 dX^2 - X^2 dX^1)^2 
+ (X^2 dX^3 - X^3 dX^2)^2 \nonumber \\
&& \quad\quad\quad\quad\quad\quad\quad\quad\quad\quad\quad\quad
\quad\quad\quad\quad + (X^3 dX^1 - X^1 dX^3)^2\Bigr) \,.
\label{SimpRiemNorm}
\eea
The curvature of a dust-filled FRW universe therefore has 
a universal structure multiplied by the (time-varying) density. 
The ratio of coefficients $2:1$ is a constant feature.

From the curvature of spacetime, one can determine $qH^2$ and $H^2 +
K$, but not $H$ and $K$ separately. However, $H$ is independently 
related to the flow of matter. Let us find the flow, calculating 
to linear order in the normal coordinates around $\Oh$. We start with the 
comoving velocity 4-vector $u^\mu = (1,0,0,0)$ and apply the 
coordinate change (\ref{tX}). We also apply the further coordinate 
change (\ref{TX}), although this has essentially no effect. In coordinates 
$X^\mu = (T,\X)$,
\bea
U^\mu &=& u^\nu \frac{\pr X^\mu}{\pr x^\nu} \nonumber \\ 
&=& \frac{\pr X^\mu}{\pr t} \nonumber \\
&=& (1, \, H X^1, \, H X^2, \, H X^3)
\label{HubFlowU}
\eea
at linear order. The 4-vector $U^\mu$ represents the Hubble 
flow, with radial velocity proportional to spatial distance from $\Oh$.

At this order we can verify the conservation of the energy-momentum 
tensor. As $\rho(t) R(t)^3$ is conserved, we have to linear order 
$\rho(t) = \rho_0 (1 - 3Ht)$, which implies that
\be
\rho(T) = \rho_0 (1 - 3HT) \,.
\label{densityevol}
\ee
The energy-momentum tensor near $\Oh$ is
\be
T^{\mu\nu} = \rho U^\mu U^\nu
\ee
with components
\bea
T^{00} &=&  \rho_0(1 - 3HT) \,, \nonumber \\
T^{0i}=T^{i0} &=&  \rho_0 (1 - 3HT) H X^i = \rho_0 H X^i \,, \nonumber \\
T^{ij} &=& \rho_0 (1 - 3HT) H X^i X^j = 0 \,,
\eea
where the final expressions retain terms only up to linear order in the
coordinates. As the Christoffel symbols now vanish at $\Oh$, the 
energy-momentum conservation law is that of Minkowski space at this 
order. We find
\bea
\pr_0 T^{00} + \pr_i T^{i0} &=& -3\rho_0 H + 3\rho_0 H =0 \,, \nonumber \\
\pr_0 T^{0j} + \pr_i T^{ij} &=& 0 \,,
\eea
so energy-momentum conservation is verified, but only at $\Oh$.
In summary, we see that $H$ is determined either from the Hubble flow of
matter, or from the time derivative of $\rho$. 

It would be interesting to extend this analysis, working with 
$T^{\mu\nu}$ at quadratic order and Christoffel symbols at linear order, to
verify energy-momentum conservation at linear order.

\vspace{7mm}

\section{Newtonian Spacetime Patches}
\vspace{4mm}

Here we show that the Newtonian approximation to spacetime structure
is valid in a spacetime patch around the point $\Oh$. By Newtonian
approximation we mean the weak field approximate solution of Einstein's
equations in the presence of static or slowly moving matter. Geodesic
motion of a test particle in this curved Newtonian spacetime patch
reproduces the test particle's Newtonian motion in flat spacetime
under the influence of the gravitational potential.

The Newtonian potential $\Phi$ is a solution of the Poisson equation
\be
\nabla^2 \Phi = 4\pi G\rho \,.
\label{Poisson}
\ee
For a spatially uniform matter density $\rho$, which may depend on time, the
solution we choose is
\be
\Phi(t, \x) = \frac{2\pi G}{3} \rho \, \x \cdot \x \,.
\label{Poissonsoln}
\ee
Here, $(t, \x)$ are time and space coordinates that vanish at $\Oh$. 
They are not the comoving coordinates that appear in eq.(\ref{FRWquad}).
The solution (\ref{Poissonsoln}) is adapted to $\Oh$ as centre, and is 
only useful in a patch around $\Oh$. Another solution would be used in the
neighbourhood of another point.

Where $\Phi$ is small, and $\rho$ slowly varying, the metric \cite{Car} 
\be
ds^2 = -(1 + 2\Phi)dt^2 + (1-2\Phi)d\x \cdot d\x
\label{Newtmet}
\ee
satisfies the linearised Einstein equations,
provided $\Phi$ satisfies eq.(\ref{Poisson}). It is essential that
$2|\Phi|$ is everywhere much less than $1$, and this limits the region
of validity of (\ref{Newtmet}). The temporal part of the metric, 
$-(1 + 2\Phi)dt^2$ is sufficient to reproduce the Newtonian equation of
motion of non-relativistic test particles,
\be
\ddot \x = -\nabla \Phi = -\frac{4\pi G}{3} \rho \, \x \,.  
\label{Newteq}
\ee
The acceleration towards the central point $\Oh$ is proportional
to the distance from $\Oh$. Eq.(\ref{Newteq}) also controls the acceleration
of the background matter, the dust. If we write $\x = R(t) \x'$, where
$\x'$ is a comoving coordinate that is time-independent, then (\ref{Newteq}) 
becomes
\be
\frac{\ddot R}{R} = -\frac{4\pi G}{3}\rho
\label{Friedeq'}
\ee
which is the Friedmann equation (\ref{Friedeq}). Moreover,
conservation of matter requires that $\rho R^3$ is constant, as in 
(\ref{masscons}). 

If we had used the solution adapted to another centre, the
acceleration would have been towards that centre. This apparent
contradiction is resolved if we can show that the different Newtonian
patches all give correct local descriptions of the
FRW universe. We return to this point in Section 4.

We want to show next that, by a change of coordinates, the
Newtonian-type metric (\ref{Newtmet}), with 
$\Phi =  \frac{2\pi G}{3} \rho \x \cdot \x$,
\be
ds^2 = -dt^2 + d\x \cdot d\x - \frac{4\pi G}{3} \rho \, \x \cdot \x \, dt^2
 - \frac{4\pi G}{3} \rho \, \x \cdot \x \, d\x \cdot d\x \,,
\label{Newtrho}
\ee
can be reduced to the Riemann normal form of the FRW metric near the point 
$\Oh$. As before, this coordinate change preserves spherical symmetry about 
$\Oh$, so the formulae are very similar to those in Section 2. 

Note that (\ref{Newtrho}) differs from flat Minkowski space only at quadratic
order, so only cubic coordinate changes are needed, with no quadratic terms.
A further simplification is that we can treat the response of the
metric to the time-varying density as instantaneous. In fact, we can
effectively ignore the time-dependence of $\rho$. We know that 
$\rho(t) = \rho_0(1 - 3Ht)$, to linear order in $t$, but the 
$t$-dependence only occurs in the metric at cubic order, beyond 
the order to which we are working. So we rewrite (\ref{Newtrho}) as 
\be
ds^2 = -dt^2 + d\x \cdot d\x - \frac{4\pi G}{3} \rho_0 \, \x \cdot \x \, dt^2
 - \frac{4\pi G}{3} \rho_0 \, \x \cdot \x \, d\x \cdot d\x \,.
\label{Newtrho0}
\ee

In (\ref{Newtrho0}) we carry out the change of coordinates
\bea
t &=& T - \frac{4\pi G}{9} \rho_0 \, T \, \X \cdot \X \,,
\nonumber \\
\x &=& \X - \frac{2\pi G}{9} \rho_0 \, T^2 \, \X 
+ \frac{2\pi G}{9} \rho_0 (\X \cdot \X) \X \,,
\label{Newtcoordchg}
\eea
and hence
\bea
dt &=& dT - \frac{4\pi G}{9} \rho_0 (\X \cdot \X \, dT 
+ 2T \, \X \cdot d\X) \,, \nonumber \\
d\x &=& d\X - \frac{2\pi G}{9} \rho_0(2T \, \X \, dT + T^2 \, d\X) \nonumber \\ 
&& + \frac{2\pi G}{9} \rho_0 (2\X (\X \cdot d\X) + (\X \cdot \X) d\X) \,.
\eea
The resulting metric is exactly the same at quadratic order as 
(\ref{SimpRiemNorm}), the Riemann form of the FRW metric near $\Oh$. 
The spacetime curvature depends on just the instantaneous density $\rho_0$.

The coordinate change (\ref{Newtcoordchg}) is easily inverted by
taking the cubic terms to the other side, and in them replacing $(T,\X)$ by 
$(t,\x)$. Then, applied to (\ref{SimpRiemNorm}) one would obtain the
metric (\ref{Newtrho0}). Therefore, by composing the coordinate
changes, one can obtain the Newtonian-type metric (\ref{Newtrho0})
starting from the FRW metric (\ref{FRWquad}), at least up to quadratic
order.  

The Newtonian-type metric doesn't determine the Hubble parameter $H$. 
As we showed at the end of section 2, a radial Hubble flow with any value of
$H$ is consistent, provided that the size of the Newtonian patch is
limited so that the velocities relative to $\Oh$ stay
non-relativistic. As before, the value of $H$ is correlated with the
time-dependence of $\rho$ in the coordinates we are using (either $t$
or $T$). At linear order, $\rho(T) = \rho_0 (1 - 3HT)$.

\vspace{7mm}

\section{Overlapping Newtonian Patches}
\vspace{4mm}

We have shown that the metric and comoving dust in an FRW universe, in
the neighbourhood of a spacetime point $\Oh$, can be described 
by a Newtonian-type metric and a radial Hubble flow. The 
spatial homogeneity of the FRW universe implies that this result must 
be true for any choice of $\Oh$. The local metric will be of identical 
form for patches centred at two distinct points with the same comoving 
time. In particular, for a pair of points with a small 
separation, whose Newtonian patches significantly overlap, the 
apparent inhomogeneity of the Newtonian-type metric
should disappear after a coordinate transformation. 

Recall the metric (\ref{Newtrho0}),
\be
ds^2 = -dt^2 + d\x \cdot d\x - \kappa \x \cdot \x \, dt^2
 - \kappa \x \cdot \x \, d\x \cdot d\x\,,
\label{Newtkap}
\ee
where we have written $\kappa = \frac{4\pi G}{3} \rho_0$. This is apparently
inhomogeneous in $\x$, as the structure changes if we set $\x = \y +
\bep$, where $\bep$ is a small constant vector, and expand around $\bep$. 
However, with a more subtle coordinate transformation, the metric
expressed in terms of $\y$ has effectively the same form as the 
original metric in terms of $\x$. This is not an exact result, but
true to first order in $\bep$ and up to a certain order in the expansion
in $\y$. 

Let us simplify the algebra by choosing $\bep = (\e,0,0)$.
The required coordinate change is of the form
\bea
t &=& u - \e\kappa y_1 u \,, \nonumber \\
x_1 &=& y_1 +\e + \half\e\kappa (y_1^2 - y_2^2 - y_3^2 - u^2) \,, \nonumber \\
x_2 &=& y_2 + \e\kappa y_1 y_2 \,, \nonumber \\
x_3 &=& y_3 + \e\kappa y_1 y_3 \,,
\label{Newtcoorchng}
\eea
which includes all types of quadratic term consistent
with the vector nature of $\bep$, and which reduces to the trivial
coordinate change when $\e = 0$. The coefficients are such that 
after the coordinate change, if one keeps the terms up to first order 
in $\e$ and up to quadratic order in $u$ and $\y$, then the metric becomes
\be
ds^2 = -du^2 + d\y \cdot d\y - \kappa \y \cdot \y \, du^2
 - \kappa \y \cdot \y \, d\y \cdot d\y \,,
\label{Newtkap'}
\ee
which has the same form as (\ref{Newtkap}).

Because the metric preserves its form to this order,
one can say that the Newtonian patch around $\Oh$ is
geometrically homogeneous, despite appearances. This gives a
novel resolution of the apparent paradox in Newtonian
cosmology that the physics picks out a choice of centre.

To further clarify why we can be satisfied with a result to this 
order, we show in the Appendix how a simpler metric, the round metric
on a 2-sphere, exhibits its homogeneity when one compares the
expansions of the metric around two closely separated points.

\vspace{7mm}

\section{Conclusions}
\vspace{4mm}

We have considered the standard, spatially homogeneous FRW universe 
filled with pressureless dust. We have shown that a patch of the 
universe, centred at any spacetime point $\Oh$, may be brought by
coordinate changes to Newtonian form, where the spacetime metric is
expressed in terms of a local Newtonian gravitational potential
$\Phi$. Our calculations have made use of the Riemann normal form of
the metric, where the Riemann curvature tensor is directly related to
the quadratic terms in the expansion of the metric around a chosen
point. The transformations from FRW to Riemann form, and from Newtonian
to Riemann form, have been given explicitly. Combining the first
transformation with the formal inverse of the second is
straightforward, and transforms the FRW metric to Newtonian form.
The agreement is at quadratic order in the expansion of the metric
around $\Oh$. This means that the explicit metrics we have presented,
(\ref{SimpRiemNorm}) and (\ref{Newtrho0}), which are truncated at 
quadratic order, osculate the FRW spacetime.

This approach, which fully uses the ideas of general relativity, and involves
fairly complicated nonlinear coordinate changes, avoids some of the
paradoxes of a purely Newtonian cosmology. For example, there is no
need to regard matter as occurring only in a large but finite
ball. Our approach also avoids the paradox that in an infinite
Newtonian cosmology one finds that the acceleration of matter appears 
to be towards a fixed centre, violating the notion of spatial homogeneity.  

We have shown how the Newtonian patches centred at nearby points are
geometrically consistent with each other. The Newtonian atlas of
patches is therefore consistent with homogeneity. Ultimately, this is
possible because the local Newtonian potential $\Phi$ is not a
geometrical invariant, but varies (by more than constant shifts) under
coordinate transformations.

It would be interesting to extend the analysis here to include a
cosmological constant, and to allow for small-scale density inhomogeneities.

\vspace{7mm}

\section*{Appendix}
\vspace{4mm}

Here we consider how the homogeneity of the round metric on the 
2-sphere is manifested when one works with expansions around 
neighbouring points.

We use a stereographic coordinate $z= z_1 + iz_2$. The
metric is
\be
ds^2 = \frac{dz d\z}{(1 + z\z)^2}
\label{spheremet}
\ee
and is invariant under SU(2) M\"obius transformations
\be
z = \frac{\alpha w + \beta}{-\bar\beta w + \bar\alpha}
\ee
with $|\alpha|^2 + |\beta|^2 = 1$. The metric has the same form in
terms of $w$ as it has in terms of $z$.

Now suppose that $\alpha = 1$ and $\beta =\e$, with 
$\e$ small and real, and work to first order in $\e$. The 
M\"obius transformation has the form
\be
z = w + \e + \e w^2 \,,
\label{Mobsmall}
\ee
so $dz = dw + 2\e w \, dw$. To first order in $\e$ and exactly in 
$w$ or $z$, one can check that the metric is formally invariant. Notice that
the transformation (\ref{Mobsmall}) is a small translation with a quadratic
correction. The patches centred at $z=0$ and $w=0$ have large
overlap. Formally, there is no problem dropping terms of higher order 
in $\e$, but they are only small provided $\e |w| \ll 1$.

The next step is the most interesting, and gets closest to the issue
of the effective homogeneity of the Newtonian-type metric. Consider the
expansion of the metric (\ref{spheremet}) around the origin, to quadratic
order in $|z|$,
\be
ds^2 = (1 - 2z\z)dz d\z \,,
\label{spherequad}
\ee
and carry out the coordinate change (\ref{Mobsmall}). One finds
\be
ds^2 = (1 -2w\w - 6\e(w \w^2 + w^2 \w))dw d\w \,.
\label{spherequady}
\ee
The form of (\ref{spherequad}) is reproduced, with $z$ replaced by 
$w$, but only if one drops the terms of first order in $\e$ and cubic in
$|w|$. This is to be expected. The approximation 
(\ref{spherequad}) has ignored terms which are quartic in $|z|$, and 
from these one obtains, using (\ref{Mobsmall}), further terms that are 
first order in $\e$ and cubic in $|w|$, which cancel those in 
(\ref{spherequady}).

The Newtonian-type metric is analogous. This has terms with coefficients
quadratic in $\x$, but there are potentially quartic terms which have been
dropped. One can only expect the metric to be effectively 
homogeneous under shifts of the origin if one drops terms that 
are simultaneously first order in the small shift parameter $\e$ and 
cubic in the coordinates $\x$. This agrees with what we found in Section 4.

Also worth noting is that if one uses the real coordinates on the
2-sphere $z_1$ and $z_2$, and analogously $w_1$ and $w_2$, then the
metric (\ref{spherequad}) is 
\be
ds^2 = (1 - 2z_1^2 - 2z_2^2)(dz_1^2 + dz_2^2) \,,
\ee 
and the coordinate change (\ref{Mobsmall}) is $z_1 
= w_1 + \e + \e(w_1^2 - w_2^2)$ and $z_2 = w_2 + 2\e w_1 w_2$. The
metric now has considerable similarity to (\ref{Newtkap}), and the
coordinate change to the second and third of (\ref{Newtcoorchng}).

\section*{Acknowledgements}

I am grateful to Gary Gibbons and George Ellis for explaining their
ideas about the Newtonian approach to cosmology, and to them and John
Barrow for general advice on FRW spacetimes.

\end{document}